\documentclass[]{aa}
\usepackage{epsfig}

\begin{document}
\title{The first 8--13 $\mu$m spectra of globular cluster red giants:
circumstellar silicate dust grains in 47\,Tucanae (NGC\,104)}
\author{Jacco Th. van Loon\inst{1},
        Iain McDonald\inst{1},
        Joana M. Oliveira\inst{1},
        A. Evans\inst{1},\\
        Martha L. Boyer\inst{2},
        Robert D. Gehrz\inst{2},
        Elisha Polomski\inst{2},
        Charles E. Woodward\inst{2}}
\institute{Astrophysics Group, School of Physical \& Geographical Sciences,
           Lennard-Jones Laboratories, Keele University, Staffordshire ST5
           5BG, UK
      \and Department of Astronomy, University of Minnesota, 116 Church Street
           SE, Minneapolis, MN 55455, USA}
\date{Received 15 November 2005; accepted 2 January 2006}
\titlerunning{Circumstellar dust in 47\,Tuc}
\authorrunning{van Loon et al.}
\abstract{We present 8--13 $\mu$m spectra of eight red giants in the globular
cluster 47\,Tucanae (NGC\,104), obtained at the European Southern Observatory
3.6m telescope. These are the first mid-infrared spectra of metal-poor,
low-mass stars. The spectrum of at least one of these, namely the extremely
red, large-amplitude variable V1, shows direct evidence of circumstellar
grains made of amorphous silicate.
}
\maketitle

\section{Introduction}

Galactic globular clusters offer us an opportunity to study the late stages of
evolution of stars with a main sequence mass $\sim0.8$ M$_\odot$ at an
accurately known distance. These clusters span a range in metallicity from
[Fe/H]$<-2$ to $\sim0$, allowing studies of the dependence on metallicity over
more than two orders of magnitude. As the most populous star clusters in the
Galaxy, they also harbour rare examples of short-lived phases in stellar
evolution.

Of particular interest for the evolution of low-mass stars and their r\^{o}le
in galactic evolution is the loss of $\sim30$ per cent of their mass during
post-Main Sequence evolution. Some of this occurs prior to the core helium
burning stage, near the tip of the first ascent Red Giant Branch (RGB), and
some on the Asymptotic Giant Branch (AGB). Alfv\'{e}n waves generated in
chromospherically active giants may drive mass loss at rates of
$\dot{M}\sim10^{-7}$ M$_\odot$ yr$^{-1}$ (Pijpers \& Habing 1989; Judge \&
Stencel 1991; Schr\"{o}der \& Cuntz 2005), but in the coolest giants a
combination of strong radial pulsations with circumstellar dust formation may
facilitate a radiation-driven wind with $\dot{M}\sim10^{-6}$ M$_\odot$
yr$^{-1}$ or more (Gehrz \& Woolf 1971; Bowen \& Willson 1991; Winters et al.\
2000). Very little is known about the dependence of dust-driven winds on
metallicity (van Loon 2006).

Infrared (IR) emission from circumstellar dust has been detected in several
globular clusters (Ramdani \& Jorissen 2001; Origlia et al.\ 2002), most
notably in 47\,Tucanae --- a massive cluster at 5 kpc (Gratton et al.\ 2003),
with [Fe/H]=$-0.66$ (Carretta \& Gratton 1997). We here present the first
mid-IR spectra of globular cluster red giants, obtained to identify the nature
of the dust grains.

\section{Observations}

\subsection{Spectroscopy}

The Thermal Infrared MultiMode Instrument (TIMMI2) on the 3.6m telescope of
the European Southern Observatory (ESO) at La Silla, Chile, was used on the
two nights of 19--21 October, 2005, to obtain low-resolution
($\lambda/\Delta\lambda\sim160$) 8--13 $\mu$m spectra of a sample of luminous
red giants in 47\,Tuc. We used a slit with a width of $1.2^{\prime\prime}$ on
all occasions except for V18+V11 which were observed simultaneously in a
$3^{\prime\prime}$ slit. The IR technique of chopping and nodding was used,
with a throw of 8--15$^{\prime\prime}$ depending on possible confusion. Total
integration times were 1 hour per target star spectrum (10 min for V2).

The target star spectra were divided through by the spectrum of a K-type
standard star taken close in time ($\sim1$ hour) and airmass (about $\pm0.1$)
to the observation of the target star. This removed most of the telluric
features, of which the ozone band around 9.5 $\mu$m is the most conspicuous.
To correct for the photospheric spectrum of the standard star, the thus
obtained quotient spectra were multiplied by template spectra representative
for the standard stars used (Cohen et al.\ 1999).

The data were reduced using the TIMMI2 pipeline and long-slit spectroscopic
procedures in the ESO Munich Image Data Analysis System (ESO-MIDAS). The
spectra were extracted by applying weights to the rows on the array parallel
to the dispersion direction, based on the signal in the row after subtraction
of a polynomial fit to the remaining sky background in the cross-dispersion
direction.

\subsection{Photometry}

We calibrated the acquisition images of the target stars against those of the
standard stars (van der Bliek, Manfroid \& Bouchet 1996; Cohen et al.\ 1999).
These were taken through the N1 filter (8.6 $\mu$m; width from 8.0 to 9.2
$\mu$m at half maximum). Aperture photometry was obtained in ESO-MIDAS using
an aperture diameter of $2.4^{\prime\prime}$. A zero point of 57.8 Jy was
adopted (van der Bliek et al.\ 1996). The resulting photometry is listed in
Table 1, where the uncertainty includes an 11 per cent calibration uncertainty
added in quadrature to the measurement error.

Conditions were good. The humidity ranged between 30--50 per cent and the air
temperature between 10--12 $^\circ$C. With an optical seeing of
0.3--1.1$^{\prime\prime}$ the mid-IR observations were always diffraction
limited ($0.6^{\prime\prime}$ at 10 $\mu$m).

\subsection{Targets}

We selected all stars in 47\,Tuc for which the possibility of IR excess
emission has been suggested on the basis of ISO imaging (Ramdani \& Jorissen
2001; Origlia et al.\ 2002), plus the two most extreme AGB stars in 47\,Tuc,
V1 and V2 (Frogel \& Elias 1988). The V numbers are according to Sawyer-Hogg
(1973), and the LW numbers are new variables found by Lebzelter \& Wood
(2005). From this sample of 11 targets we took spectra of the brightest 7
sources, together with V11 not on our target list but which could be placed in
the slit simultaneously with the observation of the target V18. The
acquisition images of the target LW19 revealed the likelihood of source
confusion in the ISO data, and no spectrum was taken.

We list in Table 1 the photometric periods, P, and 2.2 $\mu$m magnitudes,
[2.2], from Lebzelter \& Wood (2005). The [2.2] values are averages of the
observed maxima and minima based on photometry obtained from the literature,
and are thus representative of the mean brightness. It was based in part on
the work by Frogel, Persson \& Cohen (1981), who collected $JHK$ and
intermediate-band CO and H$_2$O photometry for red giants in 47\,Tuc,
including V\,1--3 that were found to be too bright to be RGB stars and
therefore must be on the AGB. These stars were also found to exhibit the
strong molecular absorption bands that are characteristic of cool, pulsating
AGB stars.

%
%
\begin{table}
\caption[]{List of targets, in order of increasing Right Ascension (J2000),
their photometric periods and 2.2 $\mu$m magnitudes from Lebzelter \& Wood
(2005), and our 8.6 $\mu$m magnitudes. Uncertain values for the pulsation
period are trailed by a colon.}
\begin{tabular}{lccccc}
\hline\hline
Nam\rlap{e}                                     &
$\alpha$ ($^{\rm h}$ $^{\rm m}$ $^{\rm s}$)     &
$\delta$ ($^\circ$ $^\prime$ $^{\prime\prime}$) &
$P$ (d)                                         &
[2.2]                                           &
[8.6]                                           \\
\hline
LW1\rlap{0}        &
00 24 02.6         &
\llap{$-$}72 05 07 &
\llap{1}21\rlap{:} &
6.40               &
$6.37\pm0.25$      \\
V26                &
00 24 07.9         &
\llap{$-$}72 04 32 &
65\rlap{:}         &
6.25\rlap{$^\dagger$} &
$6.11\pm0.22$      \\
V8                 &
00 24 08.3         &
\llap{$-$}72 03 54 &
\llap{1}55         &
6.70               &
$5.56\pm0.21$      \\
V1                 &
00 24 12.4         &
\llap{$-$}72 06 39 &
\llap{2}21         &
6.21               &
$5.19\pm0.26$      \\
V2                 &
00 24 18.4         &
\llap{$-$}72 07 59 &
\llap{2}03         &
6.29               &
$5.45\pm0.16$      \\
V18                &
00 25 09.2         &
\llap{$-$}72 02 39 &
83\rlap{:}         &
7.47               &
$7.01\pm0.48$      \\
V11                &
00 25 09.0         &
\llap{$-$}72 02 17 &
\llap{1}60\rlap{:} &
6.71               &
$6.67\pm0.29$      \\
V3                 &
00 25 15.9         &
\llap{$-$}72 03 54 &
\llap{1}92         &
6.27               &
$5.72\pm0.18$      \\
\hline
\end{tabular}\\
$^\dagger$ Blend in 2MASS; Origlia et al.\ (2002) list $K=6.55$ for V26.
\end{table}

\section{Description of the spectra}

\subsection{Silicate grains in the Mira-type variable 47\,Tuc\,V1}

%
%
\begin{figure}[tb]
\centerline{\psfig{figure=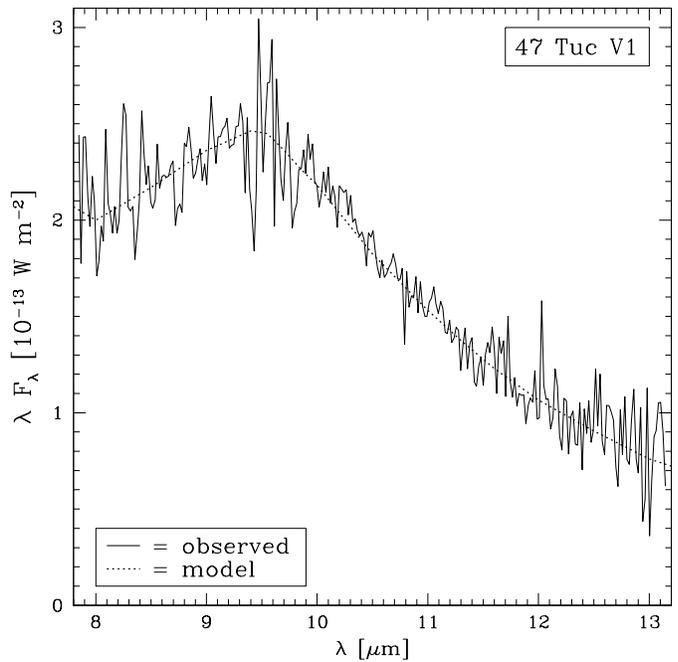,width=88mm}}
\caption[]{Observed 8--13 $\mu$m spectrum of 47\,Tuc\,V1 (solid) and a {\sc
dusty} model spectrum (dots). The broad emission around 9.5 $\mu$m is due to
amorphous silicate grains.}
\end{figure}

The reddest and most luminous variable star in 47\,Tuc, V1 shows clear
evidence for dust in its 8--13 $\mu$m spectrum (Fig.\ 1) in the form of broad
emission peaking around 9.5 $\mu$m, on top of a continuum that is
significantly less steep than that of a stellar photosphere. None of the sharp
features in the spectrum are real.

An excellent fit to the observed spectrum is obtained with the radiative
transfer model {\sc dusty} (Ivezi\'{c}, Nenkova \& Elitzur 1999), using
amorphous silicate grain properties from Draine \& Lee (1984). We assume a
grain bulk density of 3 g cm$^{-3}$ and a standard MRN grain size distribution
(Mathis, Rumpl \& Nordsieck 1977), and a stellar effective temperature of 3400
K (Lebzelter \& Wood 2005). The best fit was obtained for a dust temperature
at the inner edge of the envelope of 700 K. The radial density profile was
computed by {\sc dusty} using a radiation-driven dust wind formalism
(Ivezi\'{c} \& Elitzur 1995).

The best fit required a luminosity of 7300 L$_\odot$, and an optical depth of
0.15 at a wavelength of 0.55 $\mu$m. Assuming that the gas-to-dust ratio,
$\psi$, scales linearly with metallicity and that $\psi_\odot=200$ (van Loon,
Marshall \& Zijlstra 2005), we thus estimate a total (gas+dust) mass-loss rate
$\dot{M}=1.0\times10^{-6}$ M$_\odot$ yr$^{-1}$. The terminal wind speed is
predicted by {\sc dusty} to be 5 km s$^{-1}$.

\subsection{Mid-IR spectra of other red giants in 47\,Tuc}

%
%
\begin{figure}[tb]
\centerline{\psfig{figure=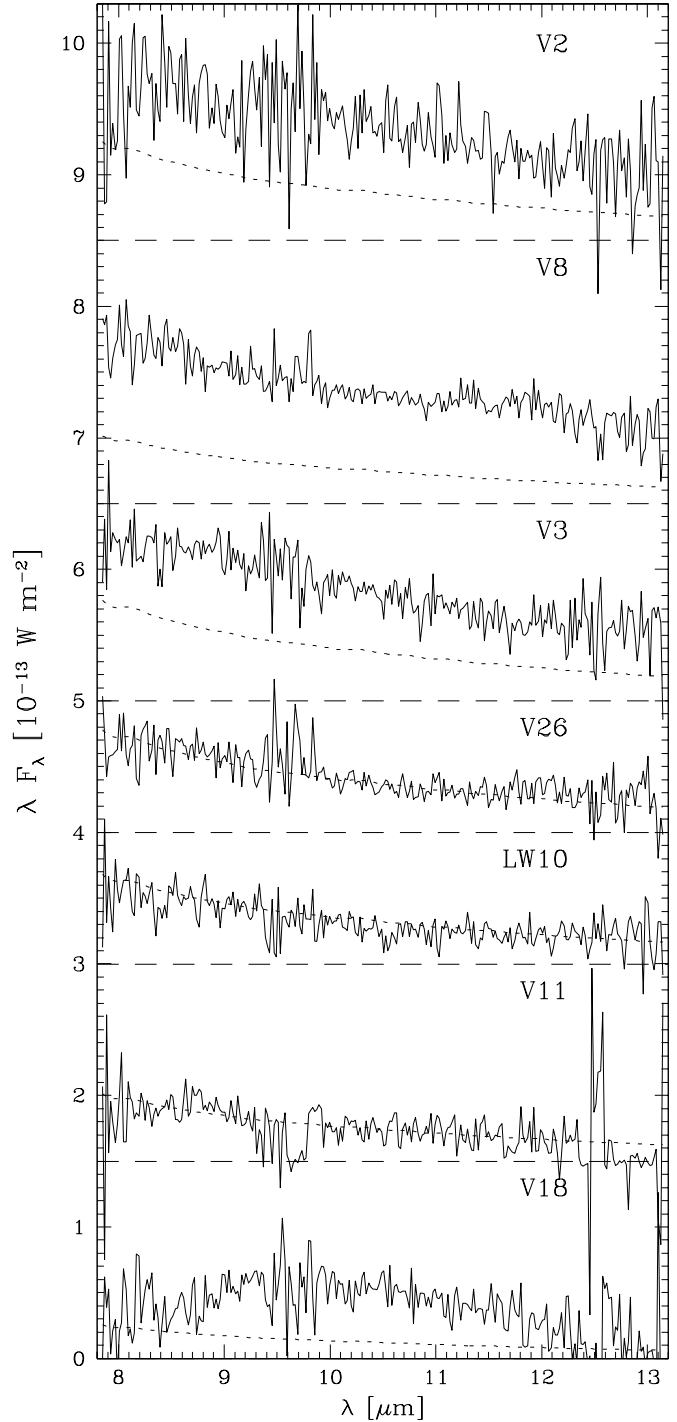,width=88mm}}
\caption[]{The 8--13 $\mu$m spectra for the other targets in 47\,Tuc, in order
(from bottom to top) of increasing 8.6 $\mu$m brightness. The dashed line
indicates the zero-level for each spectrum. The dotted curve is the template
spectrum for M3.5\,V star $\gamma$\,Crux scaled to match the 2.2 $\mu$m flux
density of the target star.}
\end{figure}

The spectra of the other targets (Fig.\ 2) are displayed together with the
template spectrum for the nearby Main Sequence star $\gamma$\,Crux of spectral
type M3.5 (a temperature similar to that of the target stars), after scaling
it to match the 2.2 $\mu$m brightness of the target stars. With $[2.2]_{\rm
2MASS}-[8.6]=-0.02$ mag, $\gamma$\,Crux shows no evidence for dust emission.
The spectra of the faint targets 47\,Tuc\,V11, LW10 and V26 follow exactly the
scaled template spectrum, and hence their spectral energy distributions can be
considered purely photospheric in origin. The faintest star in our sample at
2.2 and 8.6 $\mu$m, 47\,Tuc\,V18 shows a broad emission ``feature'' above the
photospheric continuum, without a clear peak. The brightest objects,
47\,Tuc\,V3, V8 and V2 are clearly much brighter at mid-IR wavelengths than
the photospheric template, but curiously they do not show a discrete emission
feature that could have helped identify the type of dust.

\section{Discussion}

\subsection{Red giants in 47\,Tuc which exhibit IR excess}

%
%
\begin{figure}[tb]
\centerline{\psfig{figure=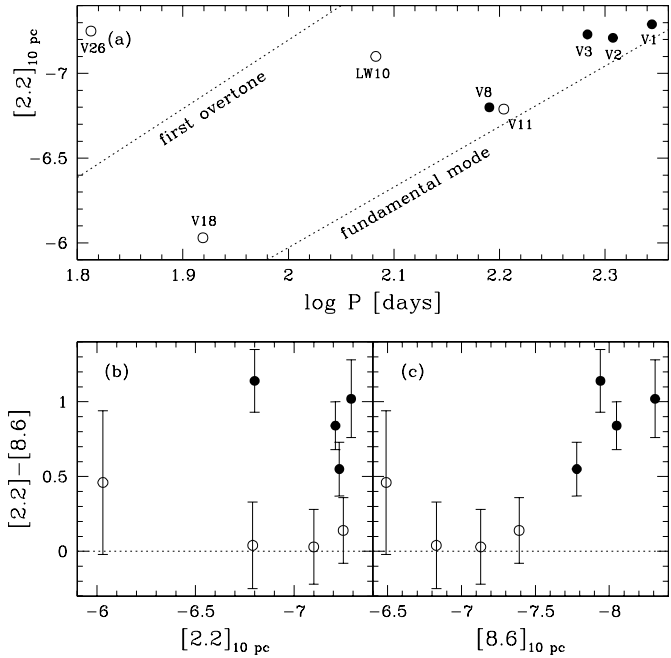,width=88mm}}
\caption[]{Absolute magnitude at 2.2 $\mu$m versus pulsation period (a), and
[2.2]--[8.6] colour versus absolute magnitude at 2.2 $\mu$m (b) and 8.6 $\mu$m
(c). Objects with [2.2]--[8.6]$>0.5$ mag have solid symbols. The two sequences
in the [2.2]$_{\rm 10 pc}$ versus P diagram are from stars in the Large
Magellanic Cloud (Ita et al.\ 2004).}
\end{figure}

The stars with the reddest [2.2]--[8.6] colours tend to have the longest
pulsation periods and highest luminosities, and straddle the sequence that in
the Large Magellanic Cloud (LMC) has been identified with stars pulsating in
the fundamental mode (Fig.\ 3a; Wood et al.\ 1999; Ita et al.\ 2004). The four
reddest stars are all Mira-type variables characterised by regular pulsations
with a large amplitude in luminosity and velocity (Lebzelter et al.\ 2005; see
also Frogel \& Whitelock (1998) and Feast, Whitelock \& Menzies 2002). The
only object without a red [2.2]--[8.6] colour that nevertheless appears to
pulsate in the fundamental mode is 47\,Tuc\,V11, whose pulsation period is
uncertain and probably shorter than the 160 days we list here (Lebzelter \&
Wood 2005).

The [2.2]--[8.6] colour correlates better with the 8.6 $\mu$m magnitude (Fig.\
3c) than with the 2.2 $\mu$m magnitude (Fig.\ 3b). Variability could have
affected the [2.2]--[8.6] colour at the level of a few 0.1 mag; the [2.2]
magnitudes are mean values but the [8.6] magnitudes are single-epoch
measurements. The amplitude at 8.6 $\mu$m is generally less than at 2.2
$\mu$m, where stars 47\,Tuc\,V\,1--3 are known to vary by about half a
magnitude (Frogel \& Elias 1988; Le Bertre 1993). The absence of negative
values for [2.2]--[8.6] in our data suggests that the red colours for
47\,Tuc\,V\,1--3 and V\,8 can not simply be explained by variability. It is
not {\it a priori} clear whether the red colours are due to excess emission
from circumstellar dust grains or simply the result of a very cool
photosphere. For comparison, the TIMMI2 standard stars (mostly K-type giants)
have $[2.2]_{\rm 2MASS}-[8.6]=-0.2$ to $+0.1$ mag.

47\,Tuc\,V18 has a reddish [2.2]--[8.6] colour still consistent with purely
photospheric emission. Ramdani \& Jorissen (2001) measured a 12 $\mu$m excess
of more than a magnitude, making it the reddest object in their sample.
Apparently, the dust emission only becomes obvious longward of 9 $\mu$m (Fig.\
2). Lebzelter et al.\ (2005) propose that this star has recently experienced a
thermal pulse.

The measurement by Ramdani \& Jorissen (2001) of the nearby 47\,Tuc\,V11 is in
perfect agreement with our measurement, showing no evidence of excess
emission. The mild excess for 47\,Tuc\,V3 in their data is also consistent
with our measurement. Of the three objects in common with Origlia et al.\
(2002) we only detect possible excess emission in 47\,Tuc\,V8, whilst the
[2.2]--[8.6] colours of 47\,Tuc\,LW10 and V26 are consistent with purely
photospheric emission despite large K--[12] colours measured by Origlia et
al.\ (2002). It is possible that some of these ISO data suffer from confusion.

\subsection{Dusty winds of metal-poor, low-mass AGB stars}

Of all members of 47\,Tuc, V1 has the longest pulsation period, largest
pulsation amplitude, coolest photosphere, and highest luminosity. It does not
therefore come as a surprise that it is this star that shows the most
convincing evidence for circumstellar dust.

Dijkstra et al.\ (2005) suggest a sequence for oxygen-rich grain mineralogy,
whereby emission from amorphous alumina (Al$_2$O$_3$) grains at 11--13 $\mu$m
(Speck et al.\ 2000) dominates the spectrum of stars with relatively low
mass-loss rates and emission from amorphous silicate takes over at higher
mass-loss rates. The reason given is that alumina has the highest condensation
temperature of all known dust grain species, at $\sim1500$ K in a typical
circumstellar environment. Silicates condense onto alumina seeds, obliterating
the signatures of the alumina in dense winds (Tielens 1990). Although the
alumina is seen at mass-loss rates of $\dot{M}\sim10^{-6}$ M$_\odot$ yr$^{-1}$
in massive AGB stars in the Large Magellanic Cloud (LMC), there is no trace of
it in the spectrum of 47\,Tuc-V1. One could speculate that the slower and more
compact wind of this metal-poor, low-luminosity star reduces the volume of the
wind that hosts bare alumina particles.

Compared to dust-enshrouded intermediate-mass AGB stars (van Loon et al.\
1999) the mass-loss rate of 47\,Tuc\,V1 is with $\dot{M}=1.0\times10^{-6}$
M$_\odot$ yr$^{-1}$ not very high, but sufficient to drive a wind by means of
radiation pressure on the dust grains (Winters et al.\ 2000). There are a few
hundred stars observed in 47\,Tuc (Alves-Brito et al.\ 2005) that are in the
last $10^8$ years of their AGB evolution (Girardi et al.\ 2000, their models
for $M=0.8$ M$_\odot$ and $Z=0.004$). This suggests that the phase in which we
now find 47\,Tuc\,V1 lasts a few $10^5$ yr, during which a few 0.1 M$_\odot$
is lost. Like in intermediate-mass AGB stars (van Loon et al.\ 2005), much of
the mass loss from metal-poor, low-mass AGB stars appears to be in the form of
a dust-driven wind. Because this phase is so brief, only very few such objects
(if any) are expected to be present in a globular cluster at any time. The
amount of gas and dust to be lost by 47\,Tuc\,V1 is very similar to that found
in the interstellar medium of the massive globular cluster M\,15 (Evans et
al.\ 2003; van Loon et al.\ 2006).

\section{Summary of conclusions}

We presented the first mid-infrared spectra of red giants in a galactic
globular cluster, 47\,Tucanae. The most evolved object, 47\,Tuc\,V1 is found
to be surrounded by dust grains made of amorphous silicate, and to experience
mass loss at a rate of $\dot{M}=1.0\times10^{-6}$ M$_\odot$ yr$^{-1}$.

\begin{acknowledgements}
We would like to thank the La Silla staff and dog for a pleasant stay at the
observatory. We thank the referee for her/his comments that helped us clarify
some points. Iain McDonald is supported by a PPARC studentship.
\end{acknowledgements}

\end{document}